\documentstyle[12pt]{article}

\catcode`\@=11
\long\def\@makefntext#1{ 
\protect\noindent \hbox to 3.2pt {\hskip-.9pt  
$^{{\ninerm\@thefnmark}}$\hfil}#1\hfill} 

\def\thefootnote{\fnsymbol{footnote}}
 \def\@makefnmark{\hbox to 0pt{$^{\@thefnmark}$\hss}}  
	
\def\ps@myheadings{\let\@mkboth\@gobbletwo
\def\@oddhead{\hbox{} 
\rightmark\hfil\ninerm\thepage}   
\def\@oddfoot{}\def\@evenhead{\ninerm\thepage\hfil 
\leftmark\hbox{}}\def\@evenfoot{}
\def\sectionmark##1{}\def\subsectionmark##1{}}

\begin{document}

\newcommand{\symbolfootnote}{\renewcommand{\thefootnote}
	{\fnsymbol{footnote}}}
\renewcommand{\thefootnote}{\fnsymbol{footnote}}
\newcommand{\alphfootnote}
	{\setcounter{footnote}{0}
	 \renewcommand{\thefootnote}{\sevenrm\alph{footnote}}}

\newcounter{sectionc}\newcounter{subsectionc}\newcounter{subsubsectionc}
\renewcommand{\section}[1] {\vspace{0.6cm}\addtocounter{sectionc}{1} 
\setcounter{subsectionc}{0}\setcounter{subsubsectionc}{0}\noindent 
	{\bf\thesectionc. #1}\par\vspace{0.4cm}}
\renewcommand{\subsection}[1] {\vspace{0.6cm}\addtocounter{subsectionc}{1} 
	\setcounter{subsubsectionc}{0}\noindent 
	{\it\thesectionc.\thesubsectionc. #1}\par\vspace{0.4cm}}
\renewcommand{\subsubsection}[1] {\vspace{0.6cm}\addtocounter{subsubsectionc}{1}
	\noindent {\rm\thesectionc.\thesubsectionc.\thesubsubsectionc. 
	#1}\par\vspace{0.4cm}}
\newcommand{\nonumsection}[1] {\vspace{0.6cm}\noindent{\bf #1}
	\par\vspace{0.4cm}}
					         
\newcounter{appendixc}
\newcounter{subappendixc}[appendixc]
\newcounter{subsubappendixc}[subappendixc]
\renewcommand{\thesubappendixc}{\Alph{appendixc}.\arabic{subappendixc}}
\renewcommand{\thesubsubappendixc}
	{\Alph{appendixc}.\arabic{subappendixc}.\arabic{subsubappendixc}}

\renewcommand{\appendix}[1] {\vspace{0.6cm}
        \refstepcounter{appendixc}
        \setcounter{figure}{0}
        \setcounter{table}{0}
        \setcounter{equation}{0}
        \renewcommand{\thefigure}{\Alph{appendixc}.\arabic{figure}}
        \renewcommand{\thetable}{\Alph{appendixc}.\arabic{table}}
        \renewcommand{\theappendixc}{\Alph{appendixc}}
        \renewcommand{\theequation}{\Alph{appendixc}.\arabic{equation}}
        \noindent{\bf Appendix \theappendixc #1}\par\vspace{0.4cm}}
\newcommand{\subappendix}[1] {\vspace{0.6cm}
        \refstepcounter{subappendixc}
        \noindent{\bf Appendix \thesubappendixc. #1}\par\vspace{0.4cm}}
\newcommand{\subsubappendix}[1] {\vspace{0.6cm}
        \refstepcounter{subsubappendixc}
        \noindent{\it Appendix \thesubsubappendixc. #1}
	\par\vspace{0.4cm}}

\def\abstracts#1{{
	\centering{\begin{minipage}{30pc}\tenrm\baselineskip=12pt\noindent
	\centerline{\tenrm ABSTRACT}\vspace{0.3cm}
	\parindent=0pt #1
	\end{minipage} }\par}} 

\newcommand{\bibit}{\it}
\newcommand{\bibbf}{\bf}
\renewenvironment{thebibliography}[1]
	{\begin{list}{\arabic{enumi}.}
	{\usecounter{enumi}\setlength{\parsep}{0pt}
\setlength{\leftmargin 1.25cm}{\rightmargin 0pt}
	 \setlength{\itemsep}{0pt} \settowidth
	{\labelwidth}{#1.}\sloppy}}{\end{list}}

\topsep=0in\parsep=0in\itemsep=0in
\parindent=1.5pc

\newcounter{itemlistc}
\newcounter{romanlistc}
\newcounter{alphlistc}
\newcounter{arabiclistc}
\newenvironment{itemlist}
    	{\setcounter{itemlistc}{0}
	 \begin{list}{$\bullet$}
	{\usecounter{itemlistc}
	 \setlength{\parsep}{0pt}
	 \setlength{\itemsep}{0pt}}}{\end{list}}

\newenvironment{romanlist}
	{\setcounter{romanlistc}{0}
	 \begin{list}{$($\roman{romanlistc}$)$}
	{\usecounter{romanlistc}
	 \setlength{\parsep}{0pt}
	 \setlength{\itemsep}{0pt}}}{\end{list}}

\newenvironment{alphlist}
	{\setcounter{alphlistc}{0}
	 \begin{list}{$($\alph{alphlistc}$)$}
	{\usecounter{alphlistc}
	 \setlength{\parsep}{0pt}
	 \setlength{\itemsep}{0pt}}}{\end{list}}

\newenvironment{arabiclist}
	{\setcounter{arabiclistc}{0}
	 \begin{list}{\arabic{arabiclistc}}
	{\usecounter{arabiclistc}
	 \setlength{\parsep}{0pt}
	 \setlength{\itemsep}{0pt}}}{\end{list}}

\newcommand{\fcaption}[1]{
        \refstepcounter{figure}
        \setbox\@tempboxa = \hbox{\tenrm Fig.~\thefigure. #1}
        \ifdim \wd\@tempboxa > 6in
           {\begin{center}
        \parbox{6in}{\tenrm\baselineskip=12pt Fig.~\thefigure. #1 }
            \end{center}}
        \else
             {\begin{center}
             {\tenrm Fig.~\thefigure. #1}
              \end{center}}
        \fi}

\newcommand{\tcaption}[1]{
        \refstepcounter{table}
        \setbox\@tempboxa = \hbox{\tenrm Table~\thetable. #1}
        \ifdim \wd\@tempboxa > 6in
           {\begin{center}
        \parbox{6in}{\tenrm\baselineskip=12pt Table~\thetable. #1 }
            \end{center}}
        \else
             {\begin{center}
             {\tenrm Table~\thetable. #1}
              \end{center}}
        \fi}

\def\@citex[#1]#2{\if@filesw\immediate\write\@auxout
	{\string\citation{#2}}\fi
\def\@citea{}\@cite{\@for\@citeb:=#2\do
	{\@citea\def\@citea{,}\@ifundefined
	{b@\@citeb}{{\bf ?}\@warning
	{Citation `\@citeb' on page \thepage \space undefined}}
	{\csname b@\@citeb\endcsname}}}{#1}}

\newif\if@cghi
\def\cite{\@cghitrue\@ifnextchar [{\@tempswatrue
	\@citex}{\@tempswafalse\@citex[]}}
\def\citelow{\@cghifalse\@ifnextchar [{\@tempswatrue
	\@citex}{\@tempswafalse\@citex[]}}
\def\@cite#1#2{{$\null^{#1}$\if@tempswa\typeout
	{IJCGA warning: optional citation argument 
	ignored: `#2'} \fi}}
\newcommand{\citeup}{\cite}

\def\fnm#1{$^{\mbox{\scriptsize #1}}$}
\def\fnt#1#2{\footnotetext{\kern-.3em
	{$^{\mbox{\sevenrm #1}}$}{#2}}}

\font\twelvebf=cmbx10 scaled\magstep 1
\font\twelverm=cmr10 scaled\magstep 1
\font\twelveit=cmti10 scaled\magstep 1
\font\elevenbfit=cmbxti10 scaled\magstephalf
\font\elevenbf=cmbx10 scaled\magstephalf
\font\elevenrm=cmr10 scaled\magstephalf
\font\elevenit=cmti10 scaled\magstephalf
\font\bfit=cmbxti10
\font\tenbf=cmbx10
\font\tenrm=cmr10
\font\tenit=cmti10
\font\ninebf=cmbx9
\font\ninerm=cmr9
\font\nineit=cmti9
\font\eightbf=cmbx8
\font\eightrm=cmr8
\font\eightit=cmti8


\centerline{
\large  Energy of a spherically symmetric charged dilaton black hole} 
\vspace{0.8cm}
\centerline{
\tenrm
 A. CHAMORRO\footnote[1]{E-mail\ :\ wtpchbea@lg.ehu.es}
 and K. S. VIRBHADRA\footnote[2]{ Present address : Tata
Institute of Fundamental Research, Homi Bhabha Road, Bombay
400005, India; E-mail\ : \ shwetketu@tifrvax.tifr.res.in} }
\baselineskip=13pt
\centerline
{\tenit 
Departamento de F\'{\i}sica Te\'{o}rica, Universidad del Pa\'{\i}s Vasco
}
\baselineskip=12pt
\centerline{\tenit 
Apartado 644, 48080 Bilbao, Spain
}
\vspace{0.9cm}
\abstracts
{
The energy associated with a static and spherically symmetric charged dilaton
black hole is obtained for arbitrary value of the coupling parameter (which 
regulates the strength of the coupling of the dilaton to the Maxwell field)
$\beta$. The energy distribution depends on $\beta$, whereas the total energy
is independent of this and is given by the mass parameter
of the black hole.
}
\vfil
\vspace{0.8cm}
\def\be{\begin{equation}}
\def\ee{\end{equation}}
\def\bea{\begin{eqnarray}}
\def\eea{\end{eqnarray}}
\def\nn{\nonumber}
\def\th{\theta}
\def\ph{\phi}
\def\lt{\left}
\def\rt{\right}

In recent years there is considerable interest in obtaining  charged 
dilaton black hole solutions and investigating their properties $[1-6]$.
Garfinkle, Horowitz, and Strominger
(GHS) $[1]$ considered the action 
\be
S\ =\ \int d^4 x \ \sqrt{-g} \ \lt[\ - R \ +\ 2 {(\nabla\Phi)}^2\ +
\ e^{-2\beta \Phi} \ F^2 \rt]
\ee
and obtained a  nice form of static and spherically symmetric charged dilaton 
black hole solution $[1,2]$, given by the line element
\be
ds^2 = B dt^2 - B^{-1}\ dr^2 - D r^2 (d\th^2 + \sin^2\th d\varphi^2),
\ee
the dilaton field $\Phi$, where 
\be
e^{2\Phi} = \lt[1-\frac{r_{-}}{r}\rt]^{(1-\sigma)/\beta},
\ee
and the  component of the  electromagnetic field tensor 
\be
F_{tr} = \frac{Q}{r^2},
\ee
where
\be
B =  \lt(1-\frac{r_{+}}{r}\rt) \lt(1-\frac{r_{-}}{r}\rt)^{\sigma},
\ee
\be
D = \lt(1-\frac{r_{-}}{r}\rt)^{1-\sigma},
\ee
and
\be
\sigma\ = \ \frac{1-{\beta}^2}{1+{\beta}^2} \ .
\ee
$r_{+}$ and $r_{-}$ are  related through 
\bea
& &2M = r_{+}\ +\ \sigma\ r_{-} ,  \nn\\
& &Q^2 \lt(1+\beta^2\rt) = r_{+}\ r_{-} \ .
\eea
$M$ and $Q$ stand for mass and charge parameters, respectively.
The surface $r = r_{+}$ is the event horizon. $\beta$ is a dimensionless free 
parameter which controls the coupling between the dilaton and the Maxwell 
fields. A change in the sign of $\beta$ is the same as a change in the sign of
the dilaton field. Therefore, it is sufficient to discuss only nonnegative 
values of $\beta$.  $\beta = 0$ in GHS solution gives the well known
Reissner-Nordstr\"{o}m (RN) solution.

It is known that  several properties of  charged dilaton black holes  depend
crucially  on the coupling  parameter $\beta$ $[2-6]$.
Recently one of the present authors and Parikh $[7]$ obtained  the energy of
a static and  spherically symmetric charged dilaton black hole for $\beta = 1$.
They found  that, similar to the case of the Schwarzschild
black hole and unlike the RN black hole, the entire energy
is confined to the interior of the black hole.
It is of  interest to investigate the energy
associated with  charged dilaton black holes for arbitrary value of 
$\beta$ to see  what  the energy distribution 
is for $\beta < 1$ as well as $\beta >1$
and  whether or not the energy is confined to  the black hole interior
for any  other value of $\beta$.

The well known energy-momentum pseudotensor of Einstein is  $[8]$
\be
\Theta_i^{\ k} =  \frac{1}{16 \pi} H^{\ kl}_{i,\ \ l} \ \ ,
\ee
where
\be
H_i^{\ kl}  = \frac{g_{in}}{\sqrt{-g}}
         \lt[-g \lt( g^{kn} g^{lm} - g^{ln} g^{km}\rt)\rt]_{,m} \ .
\ee

 Latin indices run  from $0$ to $3$. $x^0$ is the time coordinate. 
The energy and momentum components are 
\be
P_i = \frac{1}{16 \pi}
        \int \int \int {H_i^{\ 0\alpha}}_{,\alpha} \ dx^1\ dx^2\ dx^3 \ ,
\ee
where the Greek index $\alpha$ takes values from $1$ to $3$. $P_0$ 
and $P_{\alpha}$ stand, respectively, for the energy (say $E$) and  momentum components.

It is known that the energy-momentum pseudotensors,
for obtaining the energy and momentum associated with
asymptotically flat spacetimes, give the correct result if calculations are
carried out in quasi-cartesian  coordinates ( those coordinates in which the
metric $g_{ik}$ approaches the Minkowski metric $\eta_{ik}$ at large  distance
) $[8-9]$. Transforming the line element $(2)$ to 
quasi-cartesian coordinates $t,x,y,z (  x =  r \sin\th  \cos\ph,
 y =  r \sin\th \sin\ph, z =  r \cos\th ) $
one gets
\be
ds^2\ = B dt^2 - D (dx^2+dy^2+dz^2) - \frac{B^{-1}-D}{r^2} (x
dx+y dy +z dz)^2.
\ee

To obtain the energy the required components
of $ H_i^{\ kl}$ are
\bea
H_0^{\ 01} &=& \frac{2x}{r^4} 
           \lt[r\lt(\sigma r_{-} + r_{+}\rt) - \sigma r_{-} r_{+} \rt], \nn\\
H_0^{\ 02} &=& \frac{2y}{r^4} 
           \lt[r\lt(\sigma r_{-} + r_{+}\rt) - \sigma r_{-} r_{+} \rt], \nn\\
H_0^{\ 03} &=& \frac{2z}{r^4} 
           \lt[r\lt(\sigma r_{-} + r_{+}\rt) - \sigma r_{-} r_{+} \rt]. 
\eea
By using $(13)$ with $(8)$ in $(11)$, applying the Gauss theorem, and then evaluating the 
integral over the surface of a sphere of radius $r$, one gets
\be
E(r)\ =\ M\ - \ \frac{Q^2}{2r} \lt( 1 - \beta^2 \rt).
\ee

Thus one  finds that the energy distribution depends on the value of the
coupling parameter $\beta$. 
The  energy is confined to its interior  {\em {only for}} $\beta = 1$
and for all other values of 
$\beta$ the energy is shared by the  interior and exterior of the black hole.
$\beta=0$ in $(14)$ gives the energy distribution in the RN field (see also
ref. $[9]$).  $E(r)$ increases with radial distance for $\beta = 0$ 
(RN spacetime) as well as $\beta<1$, decreases for
$\beta>1$, and remains constant for $\beta = 1$.
However, the total  energy ($r$ approaching infinity in $(14)$)
is independent of $\beta$ and is given by the mass parameter of the black hole.
The details of the present contribution will be published elsewhere.
\vspace{0.2in}
\begin{flushleft}
{\bf Acknowledgements}
\end{flushleft}
This work has been partially supported by the Universidad del Pais 
Vasco under contract UPV 172.310 - EA062/93 (A.C.) and by a Basque 
Government post-doctoral fellowship (K.S.V.). We thank
A. Ach\'{u}carro, J. M. Aguirregabiria, and I. Egusquiza for discussions.
\vspace{0.2in}
\begin{flushleft}
{\bf References}\\
$[1]$ D. Garfinkle, G. T. Horowitz and A. Strominger, Phys. Rev. D43\\
\ \ \ \    (1991) 3140; Erratum : Phys. Rev. D45 (1992) 3888.\\
$[2]$ J. H. Horne and G. T. Horowitz, Phys. Rev. D46 (1992) 1340.\\
$[3]$ K. Shiraishi, Phys. Lett. A166 (1992) 298.\\
$[4]$ J. A. Harvey and A. Strominger, Quantum aspects of black holes,\\
\ \ \ \     preprint EFI-92-41, hep-th/9209055 .\\
$[5]$ T. Maki and K. Shiraishi, Class. Quant. Grav. 11 (1994) 227.\\
$[6]$ C. F. E. Holzhey and F. Wilczek, Nucl. Phys. B380 (1992) 447.\\
$[7]$  K. S. Virbhadra and J. C. Parikh, Phys. Lett. B317 (1993) 312.\\
$[8]$  C. M\o ller, Ann. Phys. (NY) 4 (1958) 347. \\
$[9]$  K. P. Tod. Proc. Roy. Soc. Lond. A388 (1983) 467;\\
\ \ \ \ \  K. S. Virbhadra, Phys. Rev. D41 (1990) 1086; Phys. Rev. D42 (1990)
 2919;\\
\ \ \ \ \  F. I. Cooperstock and S. A. Richardson, in Proc. 4th Canadian Conf.
on General\\
\ \ \ \ \ \ \ Relativity and Relativistic Astrophysics ( World Scientific, 
Singapore, 1991 );\\
\ \ \ \ \ A. Chamorro and K. S. Virbhadra, hep-th/9406148.\\

\end{flushleft}
\end{document}

\end{document}